\documentclass[aps,prl,twocolumn,showpacs,amsmath,amssymb,nofootinbib,superscriptaddress,floatfix]{revtex4}

\usepackage{bm} 
\usepackage{graphicx}
\usepackage{amssymb}
\usepackage{amsmath}
\usepackage{fancybox}
\usepackage{dcolumn} 
\usepackage[dvips]{color}
\usepackage{epic}
\usepackage{longtable}

\begin{document}

\title{
Topological delocalization of
two-dimensional 
massless
Dirac fermions 
}

\author{Kentaro Nomura}
\affiliation{
Department of Physics, Tohoku University, Sendai, 980-8578, Japan
}

\author{Mikito Koshino}
\affiliation{
Department of Physics, Tokyo Institute of Technology, 
2-12-1 Ookayama,
Meguro-ku, Tokyo 152-8551, Japan}

\author{Shinsei Ryu}
\affiliation{Kavli Institute for Theoretical Physics,
 	     University of California, 
 	     Santa Barbara, 
 	     CA 93106, 
 	     USA}

\date{\today}

\begin{abstract}
The beta function of a two-dimensional massless Dirac Hamiltonian 
subject to a random scalar potential,
which, e.g.,
underlies theoretical descriptions of graphene,
is computed numerically.
Although it belongs to, from a symmetry standpoint,
the two-dimensional symplectic class,
the beta function monotonically 
increases with decreasing conductance.
We also provide 
an argument based on the spectral flows
under twisting boundary conditions,
which shows that none of states of the massless Dirac Hamiltonian can be localized.
\end{abstract}


\maketitle

The single parameter scaling theory of Anderson localization
\cite{Abrahams79} predicts that 
the quantum transport of non-interacting disordered conductors
is characterized by the beta function
\begin{equation}
\beta(g) = 
\frac{d\ln g}{d\ln L},
\label{eq: beta function}
\end{equation}
which encodes the variation of 
the dimensionless conductance $g$ with respect to 
the system size $L$.
Once the value of the conductance 
at some length scale is known, 
the quantum transport at all length scales is constructed.
\cite{IQHE}
The property of the beta function depends on the dimensionality, 
and also on the symmetry class of the microscopic Hamiltonian,
such as spin rotation and time-reversal (TR) symmetries.\cite{Lee85}
In addition, the topological nature of wavefunctions 
also has a significant effect on quantum transport.

In this paper, we discuss
the problem of Anderson localization 
for the two-dimensional (2D)
two-component Dirac Hamiltonian
subject to a random scalar potential,
\begin{equation}
\mathcal{H}=
-{i}
\hbar v_F \boldsymbol{\sigma} \cdot 
\nabla 
+
V(\mathbf{r}).
\label{eq: single dirac hamiltonian}
\end{equation}
Here, 
$\mathbf{r}\in\mathbb{R}^2$,
$\sigma_{x,y,z}$ denote the standard Pauli matrices,
and $v_F$ the constant velocity.
The details of the random scalar potential 
$V(\mathbf{r})$ will be specified later.

The random Dirac Hamiltonian 
(\ref{eq: single dirac hamiltonian})
is of direct relevance to 
the quantum transport of disordered graphene.
\cite{Novoselov04}
Although the band structure of clean graphene
has two flavors (valleys) of two-component Dirac fermions,
the intervalley scattering 
is rather weak since spatial profile of disorder in graphene
is supposed to be smooth on an atomic scale.\cite{Shon98,Ando02}
A two-component single-flavor Dirac fermion can be
realized,
{without doubling,
on a surface of a three-dimensional 
$\mathbb{Z}_2$ topological insulator.
\cite{Moore06, Fu06a, Roy06}

The properties of the eigen functions for the ideal Dirac Hamiltonian
(Eq.\ (\ref{eq: single dirac hamiltonian}) without $V$)
are well-known:
The degeneracy point in the momentum space serves as a Dirac monopole
for the Berry connection and 
wavefunctions in the momentum space pick up 
a $\pi$ phase shift when transported around the Dirac cone. 
\cite{Ando98}

From the symmetry point of view, 
the random Hamiltonian 
(\ref{eq: single dirac hamiltonian})
belongs to the symplectic symmetry class,
as it possesses an ``effective'' TR symmetry 
\begin{equation}
{i}\sigma_y
\mathcal{H}^*
\left( -{i}\sigma_y \right)
=
\mathcal{H}.
\label{eq: TR}
\end{equation}
\cite{Ludwig94} 
The beta function of 
the 2D symplectic class
shows the weak-antilocalization 
for large $g$,
and there is a metal-insulator transition
at $g^* \sim 1.4$.
\cite{Hikami80,Asada04,Markos06}

Although being a member of the symplectic symmetry class, 
there is growing evidence that 
the beta function of the random Dirac Hamiltonian 
(\ref{eq: single dirac hamiltonian})
is qualitatively different from the
conventional one for the 2D symplectic class:
(i)
Localization of non-relativistic electrons
for strong disorder
can be understood in a picture in which 
bound states localized at potential minima
overlap with each other.
However, a Dirac fermion cannot be trapped by a potential well
irrespective of the well depth \cite{Dong98,Katsnelson06},
and hence is naively expected to have a strong tendency
not to be insulating.
This makes a physical picture for 
the strongly disordered regime of the Dirac fermions different from the conventional case,
although physics of Anderson localization 
cannot fully be understood in terms of potential trapping.
(ii) As observed by Ando \textit{et al.}, 
the Berry phase $\pi$ that is accumulated around the Dirac cone 
in the momentum space leads to a 
destructive interference between a  
back scattering process and its TR counterpart,
leading to the complete absence of back scattering.
\cite{Ando98}
(iii)
The non-linear sigma model 
(NL$\sigma$M, a field theory for diffusion modes)
for the random Dirac Hamiltonian
(\ref{eq: single dirac hamiltonian})
has a $\mathbb{Z}_2$ topological term.
\cite{Fendley01,Ostrovsky07,Ryu07}
It has a little effect in the metallic regime, but should change 
the renormalization group flow in the strongly disordered regime.
Ostrovsky \textit{et al.} \cite{Ostrovsky07} conjectured
the NL$\sigma$M with the topological term has
three fixed points 
(metallic fixed point, 
 metal-semi-metal transition, and 
 semi-metal attractive fixed point).
(iv) There are numerical studies that indicate 
the increase of the conductance with system size 
even for $g \lesssim 1.4$.
\cite{Nomura07,Rycerz06}

The purpose of this paper is to compute the beta function of
(\ref{eq: single dirac hamiltonian}) numerically, 
and compare it with
conventional system of the 2D symplectic class.  
We find that the beta function of the Dirac model 
is always larger than or equal to zero, 
showing that all states are delocalized even in the strong disorder
regime, in contrast to the conventional case.  
We also provide a spectral-flow argument that clearly shows that the
localization of Dirac fermions is forbidden. 

\begin{figure}
  \begin{center}
 \includegraphics[width=6cm,clip]{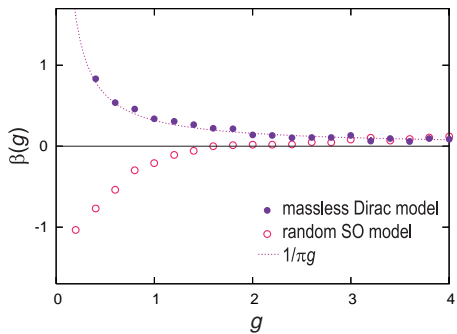} 
\caption{
\label{fig: beta.eps}
(Color online)
The beta functions of the random Dirac Hamiltonian
(\ref{eq: single dirac hamiltonian})
(closed circles)
and the random spin-orbit model
(\ref{eq: random SO model})
(open circles).
The broken line represents the one-loop beta function
of the conventional 2D symplectic class,
$\beta(g)\sim 1/\pi g$.
\cite{Hikami80}
}
\end{center}
\end{figure}

We compute the diagonal conductance (conductivity) of 
the random Dirac Hamiltonian (\ref{eq: single dirac hamiltonian})
by evaluating the Kubo formula
\begin{equation}
 g =-\frac{i 2\pi \hbar^2 }{L^2}
\sum_{n,n'}\frac{f(E_{n})-f(E_{n'})}{E_{n}-E_{n'}}
\frac{\langle n|v_x|n'\rangle\langle n'|v_x|n\rangle}{E_{n}-E_{n'}+ i\eta},
\label{kubo}
\end{equation}
where
$\mathbf{v}$ is the velocity operator, 
${\bf v}=i[\mathcal{H},{\bf r}]/\hbar
=v_F {\mbox{\boldmath$\sigma$}}$, 
$f(E)$ is the Fermi-Dirac function at zero-temperature,
$\eta$ in the energy denominator is a smearing factor,
and 
$|n\rangle$ denotes an eigenstate with energy $E_n$
of the Dirac equation 
in the presence of the random potential.

The massless two-component Dirac equation 
cannot be regularized, without breaking the TR symmetry, 
by putting the system on a lattice.
We thus work in the momentum space by
introducing 
a hard cutoff at a sufficiently large momentum $\Lambda$. 
The eigenstates $\{| n\rangle\}$ and energies $\{E_n\}$ are 
then obtained by numerically diagonalizing the Dirac Hamiltonian 
with disorder in the momentum-pseudospin basis.
Typically, we take $\Lambda \sim 20 \times 2\pi /L$,
where about $2000$ $k$-points are included.
We assume that the disorder potential is sufficiently weak
so that the level broadening caused by disorder around 
the Dirac point
is much smaller than the cut-off energy $\propto \Lambda$.

The smearing factor $\eta$ in the denominator of Eq.\ (\ref{kubo}) 
accounts for the finite switch-on time of the electric field required
for a dissipative current response.  
Physical arguments suggest that
 $\eta$ has to be at least as large as
$\hbar/ T_L$ where $T_L$ 
is the escape time from the system of interest.
The escape time can be estimated 
from the Thouless energy $\langle \Delta E\rangle$
by the uncertainty relation 
$\langle \Delta E\rangle T_L\simeq \hbar$,
where $\Delta E$ is the eigenvalue difference
between periodic and antiperiodic boundary
conditions and $\langle\,\rangle$ is
the geometric mean over disorder realizations.
\cite{Ando83,Thouless81}
Indeed, it is 
reported in Ref.\ \cite{Nomura07}
that $g$ is reasonably insensitive to $\eta$
when $\eta\simeq\langle \Delta E\rangle$.

We assume that
the scalar potential disorder $V(\mathbf{r})$ 
is generated by 
randomly distributed impurities centered at $\mathbf{R}_I$,
each of which contributes to $V(\mathbf{r})$
with a scattering potential $U({\bf r}-{\bf R}_I)$,
\begin{equation}
 V({\bf r})=\sum_{I=1}^{N_i}U({\bf r}-{\bf R}_I).
\end{equation}
We considered two types of scattering potentials
 $U({\bf r})$:
the Gaussian correlated potential,
$U({\bf q})=u\exp(-q^2l_0^2/2)$,
and 
the Thomas-Fermi potential,
$U({\bf q})=u/(q+l_0^{-1})$,
where $U(\mathbf{q})$ is the Fourier transform
of $U(\mathbf{r})$, $u$ represents the disorder strength,
and $l_0$ the range of the potential.
Typically 5000 disorder configurations were used for averaging. 
The conductance was calculated 
for various sets of parameters,
$N_i$, $u$, $l_0$ and filling ($E_F$);
The number of scatterers $N_i$ was 1-10 times as large as 
the maximum number of carriers at each size;
The range of the potential 
was changed upto 1/30 of the minimal system size.

We note that typical length scales are hardly determined 
from naive considerations at the Dirac point 
($E_F^{-1}\rightarrow \infty$). 
Indeed 
the mean free path 
at the Dirac point,
estimated by the golden rule,
diverges for uncorrelated short-range scattering 
($l_0 \to 0 $)\cite{Shon98}, 
while it vanishes
for long-range Coulomb scattering
($l_0 \to \infty$)\cite{Nomura07}.
Nevertheless we do not need the specific length scale
since the beta function is defined 
as a logarithmic derivative in Eq.\ (\ref{eq: beta function}).

To compare our results with the conventional 2D
symplectic class,
we compute, by the same method,
the beta function of 
the random spin-orbit (SO) coupling model given by
\begin{eqnarray}
\mathcal{H} &=&({-i\hbar\nabla})^2/2m+V({\bf r})+V_{\rm so},
\nonumber \\
V_{\rm so} &=& -\frac{1}{2}\{\lambda({\bf r}),-i{\nabla}\}\times
 {\mbox{\boldmath$\sigma$}}\cdot{\hat {\bf z}}.
\label{eq: random SO model}
\end{eqnarray}
Note that the velocity operator in this model is spin-dependent.
We assume an uncorrelated short-range distribution 
for $\lambda({\bf r})$ and $V({\bf r})$.

Figure\ 1 shows the beta function of the Dirac model (filled circles)
and 
the random SO coupling model
(open circles).
The latter agrees with
the known behavior of the beta function 
of the 2D symplectic universality class:
there is a metallic phase with weak anti-localization effect 
($\beta(g)\sim 1/\pi g$)
when $g$ is large, whereas there is a localized phase for small $g$;
there is a metal-insulator transition at
$g^* \sim 1.5$ that separates the two phases.

For large $g$, the beta function of the random Dirac model
behaves similarly as that of the random SO model;
this is as expected 
since when $g$ is large, 
the $\mathbb{Z}_2$ topological effect 
in the NL$\sigma$M is small.

We observe that the single parameter scaling holds reasonably well in
both models as shown in Fig.\ 1. \cite{Bardarson07}
In a sharp contrast to the conventional case,
the beta function of the Dirac model monotonically 
increases with decreasing $g$ well below $g^*\sim 1.4$: 
A 2D massless Dirac fermion cannot be localized by 
a random scalar potential.
\cite{footnote: phase diagram}
The numerical beta
function of the Dirac model is well fitted by
the one-loop beta function of the symplectic class even in the strongly disordered regime $g\lesssim 1$.

\begin{figure}
  \begin{center}
  \includegraphics[width=6cm,clip]{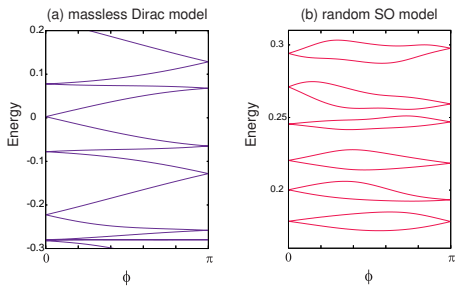} 
\caption{
\label{fig: ephi.eps}
(Color online)
The evolution of energy spectra as a function of 
the twist angle $\phi$
for 
the random massless Dirac model (left)
and
the random spin-orbit model (right).
}
\end{center}
\end{figure}

The absence of localization in the Dirac model
can  intuitively be understood by examining 
the spectral flow 
induced by twisting boundary conditions.
Let us consider a finite and disordered system
described by Eq.\ (\ref{eq: single dirac hamiltonian}),
and impose the boundary conditions in both $x$ and $y$ directions,
with phase factors $\exp(i\phi_x)$ and $\exp(i\phi_y)$, respectively.
For simplicity we set $\phi_y = 0$
and discuss the energy levels as a function of $\phi_x\equiv\phi$.
The TR symmetry holds at $\phi=0,\pi$,
where $\exp(i\phi)$ is real,
leading to the Kramers degeneracy.
We assume the cutoff to be infinity
(the effects of the finite cutoff will be discussed later).
Fig.\ \ref{fig: ephi.eps}(a) shows an example
of spectral flow
obtained for the 2D Dirac model with a specific disorder configuration.
An essential observation is that
Kramers pairs always change their partners 
as the energy spectrum evolves from $\phi=0$ to $\pi$;
if the energy eigenvalues $\{E_n \}$ are paired as
$\cdots, (E_{n},E_{n+1}), (E_{n+2},E_{n+3}),\cdots$
at $\phi=0$,
then they are paired as
$\cdots, (E_{n-1},E_{n}), (E_{n+1},E_{n+2}),\cdots$
at $\phi=\pi$.
Here eigenvalues $E_n$ are ordered in ascending order
\cite{Fu06}.
In contrast, the non-relativistic electron system with 
SO coupling has a different type of the `band-line topology'
as shown in Fig.\ \ref{fig: ephi.eps}(b);
energy eigenvalues do not change their partners 
as the spectrum evolves from $\phi=0$ to $\pi$.

We can find the origin of this topological structure in the ideal spectrum.
In the absence of disorder,
the Dirac model has a set of eigenvalues
$E_{n_x,n_y,s}(\phi) = (2\pi/L)\hbar v_F s \left[(n_x + \phi)^2 + n_y^2\right]^{1/2}$,
where $s=\pm 1$ and $n_x, n_y \in \mathbb{Z}$.
For example, two degenerate states at $\phi = 0$ with 
zero energy ($s=\pm 1$ and $n_x=n_y=0$)
become apart as $\phi$ increases
and never stick together; 
each couples with other partners at $\phi=\pi$.
As we introduce disorder, 
energy eigenvalues move around but
the way eigenvalues are paired between $\phi=0$ and $\pi$
can never be altered,
since each Kramers doublet remains sticked at $\phi=0$ and $\pi$.
In other words, it is impossible to change the 
topology of the `band-line' continuously 
from the type of Fig.\ \ref{fig: ephi.eps}(a) to (b),
without breaking the TR symmetry.

If a state is exponentially localized,
its eigen energy must be insensitive to 
the boundary phase factor,
i.e., the `band width' of $E_n(\phi)$ is exponentially small
compared with the average level spacing 
\cite{Thouless81}.
In the Dirac model, however, it is impossible
because all the band lines are connected through 
the Kramers doublets at $\phi=0,\pi$
so that the band width cannot be smaller than the level spacing.
We thus conclude that there are no localized states
in the Hamiltonian (\ref{eq: single dirac hamiltonian}).
In the non-relativistic electron system, in contrast,
the structure of the spectrum in Fig.\ \ref{fig: ephi.eps} (b)
does not, at least,  prohibit localization,
and states indeed tend to be localized 
for strong disorder.

We note that, in order for the above argument to be valid,
we have to assume that the energy band 
continues from $-\infty$ to $\infty$.
Indeed, if we have a finite cutoff, 
the TR symmetry must be broken either
at $\phi=0$ or $\pi$.
Although this may alter the band-line topology 
around the band edges,
the low-energy states around the Dirac point are
hardly affected as long as the disorder potential 
is long-ranged
and the cut-off is large enough.
With increasing the disorder strength, one would naively
expect Anderson localization first takes place
at band edges (cut-off) 
and the Dirac point (can be viewed as a point at which two band edges meet
accidentally). 
The former goes away as we send the cutoff
to infinity, while the latter is 
protected from localization
by the topology of the spectral flow.

Although the honeycomb lattice system involves 
a coupling of the two valleys (flavors),
a similar delocalization effect 
should manifest itself 
when intervalley scattering is negligibly weak. 
On the other hand, when atomic-scale scatterers dominate,
the intervalley scattering randomizes the Berry phase 
and the nature of interference is changed
to enhance localization.\cite{Ando98}
In the Dirac band, the inter-valley scattering time
depends on the Fermi energy as $\propto 1/|E_F|$\cite{Shon98}, 
and thus is more important in the highly doped regime.

The present calculation
suggests that the Dirac fermion system exhibits the
positive magnetoresistance.
On the other hand, the recent graphene experiments 
indicate somewhat complicated situations:
A magnetoresistance study \cite{Wu07} clarified that 
highly doped epitaxial graphene exhibits 
a crossover between positive and negative magnetoresistance 
induced by changing the temperature
as expected theoretically in Refs.\,\cite{Ando02,McCann06}.
For isolated single graphene sheets\cite{Novoselov04},  
however,
experiments show that (i) the conductivity hardly changes 
in a wide range of temperature near the Dirac point, 
while (ii) the magnetoresistance is weakly positive
at low carrier densities.\cite{Morozov06}
Although a number of theoretical scenarios have been proposed, 
including 
effects of microscopic ripples
\cite{Morozov06,Morpurgo06}, 
trigonal warping terms
\cite{McCann06}, 
and edges
\cite{Louis07}, 
there is no consensus at this moment.
Taking into account these effects 
in addition to the random scalar potential in 
Eq.\ (\ref{eq: single dirac hamiltonian})
will be done elsewhere.

Although our focus in this paper is on 2D,
the argument based on the topology of
the spectral flow applies equally well to
the 1 and 3D two-component massless Dirac fermion
with the effective TR symmetry:
A two-component massless Dirac fermion cannot be localized 
by a random scalar potential in all 1, 2, and 3D.\cite{Abanov00}
Since $d$-dimensional two-component massless Dirac fermion
can be viewed as a gapless boundary mode
of $\mathbb{Z}_2$ topological insulators 
in $(d+1)$D ($d=1,2,3$)\cite{Moore06, Fu06a, Roy06},
our discussion above concludes that 
a surface of a (strong) $\mathbb{Z}_2$ topological
insulator is always metallic, robust against disorder.
This is consistent with the speculation 
in Ref.\ \cite{Fu06a} in the context of quantum spin Hall effect.
Such metallic surface states can be called
a ``topological metal'' \cite{Fu06a}.

After completion of this work,
we became aware of a similar numerical result at the Dirac point,
obtained independently in Ref.\ \cite{Bardarson07}.

The Authors acknowledge helpful interactions with 
A. H. Castro-Neto, L. Fu,
A. Geim, 
A. W. W. Ludwig, 
A. H. MacDoanld, 
and D. N. Sheng.
K.\ N.\ and S.\ R.\  acknowledge the workshop
``Electronic Properties of Graphene'' 
at the Kavli Institute for 
Theoretical Physics at Santa Barbara, where this work was initiated.
This work was supported 
by the National Science Foundation
under Grant No.\ PHY05-51164.


\begin{thebibliography}{10}

\bibitem{Abrahams79}
E.\ Abrahams, P.\ W.\ Anderson, D.\ C.\ Licciardello, and T.\ V.\ Ramakrishnan,
Phys.\ Rev.\ Lett.\ \textbf{42}, 673 (1979).

\bibitem{IQHE}
An exception to the single parameter scaling 
is the two parameter scaling of the
integer quantum Hall effect:
D.\ E.\ Khmelnitskii,
JETP Lett. \textbf{38}, 552 (1983);
A.\ M.\ M.\ Pruisken,
Nucl.\ Phys.\ B \textbf{235}, 277 (1984).


\bibitem{Lee85}
Patrick A. Lee and T. V. Ramakrishnan, 
Rev. Mod. Phys. \textbf{57}, 287 (1985).

\bibitem{Novoselov04}
K.\ S.\ Novoselov,
A.\ K.\ Geim,
S.\ V.\ Morozov,
D.\ Jiang,
Y.\ Zhang,
S.\ V.\ Dubonos,
I.\ V.\ Grigorieva, and
A.\ A.\ Firsov,
Science \textbf{306}, 666 (2004).


\bibitem{Shon98}
H. H. Shon and T. Ando, 
J. Phys. Soc. Jpn. {\bf 67}, 2421 (1998).

\bibitem{Ando02}
T.\ Ando and H.\ Suzuura,
J.\ Phys.\ Soc.\ Jpn.\ \textbf{71},
2753 (2002);
H.\ Suzuura and T.\ Ando,
Phys.\ Rev.\ Lett.\ \textbf{89}, 266603 (2002).


\bibitem{Fu06a}
L.\ Fu, C.\ L.\ Kane, and  E.\ J.\ Mele,
Phys.\ Rev.\ Lett.\ \textbf{98}, 106803 (2007);
L.\ Fu and C.\ L.\ Kane,
Phys. Rev. B \textbf{76}, 045302 (2007).

\bibitem{Moore06}
J. E. Moore and L. Balents,
Phys.\ Rev.\ B \textbf{75}, 121306(R) (2007). 

\bibitem{Roy06}
R.\ Roy,
\texttt{cond-mat/0607531} (unpublished).

\bibitem{Ando98}
T. Ando and T. Nakanishi,  
J. Phys. Soc. Jpn.  \textbf{67}, 1704 (1998);
T. Ando, T. Nakanishi, and R. Saito, \textit{ibid},
2857 (1998).


\bibitem{Ludwig94} 
A.\ W.\ W.\ Ludwig, M.\ P.\ A.\ Fisher, R.\ Shankar, and G.\ Grinstein, 
Phys.\ Rev.\ B \textbf{50}, 7526 (1994).

\bibitem{Hikami80}
S.\  Hikami, A.\ I.\ Larkin, and Y.\ Nagaoka,
Prog.\ Theor.\ Phys.\ \textbf{63}, 707 (1980).

\bibitem{Asada04}
Y.\ Asada, K.\ Slevin, and T.\ Ohtsuki,
Phys. Rev. B \textbf{70}, 035115 (2004).

\bibitem{Markos06}
P. Markos and L. Schweitzer,
J.\ Phys.\ A: Math.\ Gen.\ \textbf{39}, 3221 (2006).



\bibitem{Dong98}
S.-H. Dong, X.-W. Hou, and Z.-Q. Ma, 
Phys. Rev. A \textbf{58}, 2160 (1998).

\bibitem{Katsnelson06}
M. I. Katsnelson, K. S. Novoselov, and A. K. Geim, 
Nature Phys. {\bf 2}, 620 (2006).

\bibitem{Fendley01}
P.\ Fendley, 
Phys.\ Rev.\ B \textbf{63}, 104429 (2001).


\bibitem{Ostrovsky07}
P.\ M.\ Ostrovsky, I.\ V.\ Gornyi, and A.\ D.\ Mirlin,
Phys. Rev. Lett. \textbf{98}, 256801 (2007).

\bibitem{Ryu07}
S.\ Ryu, C.\ Mudry, H.\ Obuse, and A.\ Furusaki,
\texttt{cond-mat/0702529}  (unpublished).

\bibitem{Nomura07}
K. Nomura and A. H. MacDonald,
Phys.\ Rev.\ Lett.\ \textbf{98}, 076602 (2007).

\bibitem{Rycerz06}
 A.\ Rycerz, 
J.\ Tworzyd{\l}o,
 and C.\ W.\ J.\ Beenakker, 
\texttt{cond-mat/0612446}  (unpublished).



\bibitem{Ando83}
T. Ando, 
J. Phys. Soc. Jpn. {\bf 52}, 1740 (1983); 
{\bf 53} 3101 (1984); {\bf 53} 3126 (1984).

\bibitem{Thouless81} 
D. J. Thouless and S. Kirkpatrick, 
J. Phys. C  \textbf{14}, 235 (1981);
Y. Imry, {\it Introduction to Mesoscopic Physics}, 
Oxford University Press (1997).







\bibitem{Bardarson07}
For a careful discussion on 
the single parameter scaling at the Dirac point,
see
J.\ H.\ Bardarson,
J.\ Tworzyd{\l}o,
P.\ W.\ Brouwer,
and C.\ W.\ J.\ Beenakker,
\texttt{arXiv:0705.0886} (unpublished).






\bibitem{footnote: phase diagram}
Our numerical beta function indicates that 
the phase diagram obtained for
the disordered quantum spin Hall system
by
H.\ Obuse, A.\ Furusaki, S.\ Ryu, and C.\ Mudry,
Phys. Rev. B \textbf{76}, 075301 (2007),
and by
A.\ M.\ Essin and J.\ E.\ Moore,
\texttt{arXiv:0705.0172}  (unpublished),
is a generic one.
In other word, whenever two different
2D $\mathbb{Z}_2$ insulators are connected by a 
insulator-insulator transition,
it should be described by the universality
class of the integer quantum Hall effect.


\bibitem{Fu06}
A similar switching of Kramers pairs
was found in the position of the Wannier states
in the $\mathbb{Z}_2$ pumping,
L.\ Fu and C.\ L.\ Kane,
Phys. Rev. B \textbf{74}, 195312 (2006).





\bibitem{Wu07}
Xiaosong Wu, Xuebin Li, Zhimin Song, Claire Berger, and Walt A. de Heer,
Phys. Rev. Lett. \textbf{98}, 136801 (2007).


\bibitem{McCann06}
E. McCann, K. Kechedzhi, Vladimir I. Falko, H. Suzuura, T. Ando, and B. L. Altshuler, 
Phys. Rev. Lett. \textbf{97}, 146805 (2006).

\bibitem{Morozov06}
S. V. Morozov, K. S. Novoselov, M. I. Katsnelson, F. Schedin, L. A. Ponomarenko, D. Jiang, and A. K. Geim, 
Phys. Rev. Lett. \textbf{97}, 016801 (2006).


\bibitem{Morpurgo06}
A. F. Morpurgo and F. Guinea, 
Phys. Rev. Lett. \textbf{97}, 196804 (2006).



\bibitem{Louis07}
E. Louis, J. A. Verges, F. Guinea, and G. Chiappe ,
Phys. Rev. B \textbf{75}, 113407 (2007).


\bibitem{Abanov00}
From the field theory point of view,
the topological term in the NL$\sigma$M
is responsible for the delocalization
in 1 and 2D.
The 3D NL$\sigma$M for 
the symplectic symmetry class does not allow
topological terms, but 
a Chern-Simons type term is possible, 
which might be responsible 
for the delocalization. See for example,
A.\ G.\ Abanov,
Phys.\ Lett.\ B \textbf{492}, 321 (2000).
     



\end{thebibliography}
\end{document}